\documentclass[a4paper, oneside, twocolumn, notitlepage, 10pt]{extarticle_ecoc}
\usepackage{ecoc}

\begin{document}
\selectlanguage{english}    


\title{Experimental validation of the closed-form GN model accounting for distributed Raman amplification in an S+C+L-band hybrid amplified long-haul transmission system}%

\author{
    Jiaqian~Yang\textsuperscript{(1)}, Henrique~Buglia\textsuperscript{(1)}, Eric~Sillekens\textsuperscript{(1)}, Mingming~Tan\textsuperscript{(2)}, Pratim~Hazarika\textsuperscript{(2,3)},\\ Dini~Pratiwi\textsuperscript{(2)}, Ronit~Sohanpal\textsuperscript{(1)}, Mindaugas~Jarmolovičius\textsuperscript{(1)}, Romulo~Aparecido\textsuperscript{(1)}, \\Ralf~Stolte\textsuperscript{(4)}, Wladek~Forysiak\textsuperscript{(2)},  Polina~Bayvel\textsuperscript{(1)}, Robert~I.~Killey\textsuperscript{(1)}
}

\maketitle                  


\begin{strip}
 \begin{author_descr}
 
   \textsuperscript{(1)} Optical Networks Group, UCL (University College London), London, UK, 
   \textcolor{blue}{\uline{jiaqian.yang.18@ucl.ac.uk}}\\
   \textsuperscript{(2)} Aston Institute of Photonic Technologies, Aston University, Birmingham, B4 7ET, UK\\
   \textsuperscript{(3)} Corning Optical Communications, St David’s Park, Ewloe, CH5 3XD, UK\\
   \textsuperscript{(4)} Coherent / Finisar, New South Wales, Australia

 \end{author_descr}
\end{strip}

\setstretch{1.1}
\renewcommand\footnotemark{}
\renewcommand\footnoterule{}


\begin{strip}
  \begin{ecoc_abstract}
    The accuracy of a recently-developed closed-form GN nonlinear interference model is evaluated in experimental 1065~km S+C+L band WDM transmission with backward Raman pumping. The model accurately estimates the nonlinear interference and ASE with total SNR error of less than 0.6~dB. \textcopyright2024 The Author(s)
  \end{ecoc_abstract}
\end{strip}

\setlength{\abovedisplayskip}{3pt}
\setlength{\belowdisplayskip}{3pt}

\section{Introduction}
Ultra-wideband (UWB) wavelength-division multiplexed (WDM) signal transmission has enabled significant capacity growth over standard single-mode fibre (SSMF) in both single-span and long-haul transmissions~\cite{puttnam2024_264,elson2024continuous,puttnam2024_402,hamaoka2024_110}, leading to an urgent need for maximising throughput by 
using the closed-form Gaussian noise (GN) model~\cite{semrau2019closed,poggiolini2022closed}. With the increasing use of distributed Raman amplification (DRA) to enhance throughput in multi-band long-haul systems, the estimation of transmission performance becomes challenging due to the need to calculate nonlinear interference (NLI) noise in the presence of hybrid (lumped+Raman) amplification, inter-channel stimulated Raman scattering (ISRS), and amplified spontaneous emission (ASE) noise from the Raman coupled differential equations~\cite{buglia2024throughput}. 

Some recent experimental investigations of integral GN model~\cite{semrau2018gaussian}-based signal-to-noise ratio (SNR) estimation include a hybrid distributed Raman-erbium-doped fibre amplifier (EDFA) amplified 630~km link ~\cite{galdino2019study} by numerically solving the Raman equations, covering 91~nm optical bandwidth in C+L-band, as well as  C+L-band long-haul transmission using an integral-based GN model~\cite{zefreh2020characterization} solving the coupled Raman equations~\cite{kimura2024accurate}, taking into account the depletion of Raman pumps. However, in the latter, the system performance was dominated by back-to-back (B2B) and ASE noise and the study could not verify the accuracy of the NLI noise estimation through the use of the GN model. 

In this paper, we experimentally assessed the accuracy of a recently-proposed closed-form GN model accounting for both the inter-channel stimulated Raman scattering effect and arbitrary signal power profiles due to DRA~\cite{buglia2023closedraman}. This model, which utilises a semi-analytical solution to obtain the signal profile in the presence of the DRA and ISRS, was used to estimate the 12.4~THz S+C+L-band hybrid Raman-EDFA-thulium-doped fibre amplifier (TDFA) amplified long-haul transmission performance in the presence of backward propagating Raman pumps with total power of up to 1.5~W. In the experiment, the transceiver noise was minimised by using a relatively low symbol rate of 32~GBaud per WDM channel, to highlight the impact of accumulated ASE and NLI noise on the overall performance in the recirculating loop transmission. It is shown that the model gives an accurate SNR estimation, with a total SNR error of less than 0.38~dB and 0.60~dB over 355~km and 1065~km SSMF, respectively.
\vspace{-10pt}

\section{Experimental Demonstration}
\begin{figure*}
    \centering
    \includegraphics[width=\linewidth]{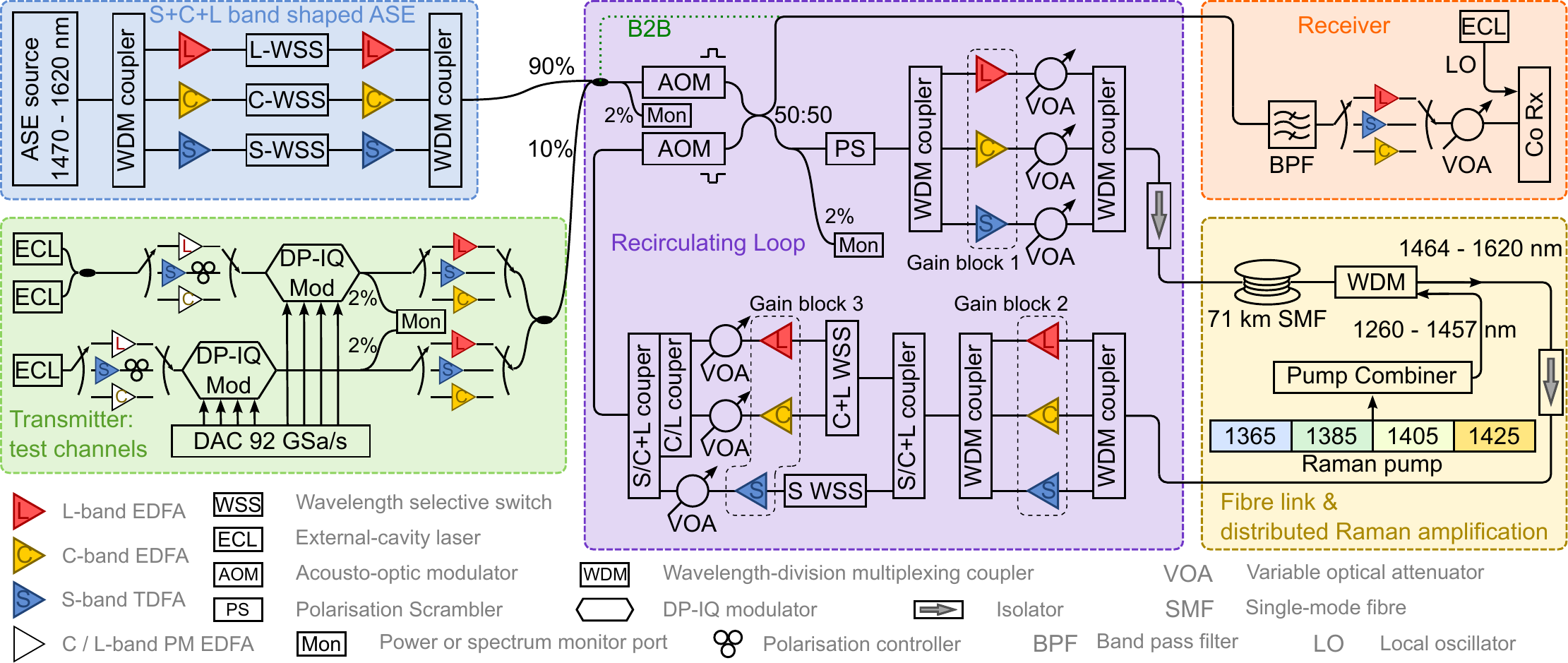}
    \caption{Experimental setup for the S+C+L-band recirculating loop transmission with hybrid Raman-EDFA-TDFA amplification.}
    \label{fig:exp}
\end{figure*}
As shown in Fig.~\ref{fig:exp}, three carriers were generated by tunable external-cavity lasers, and amplified by polarisation-maintaining EDFAs or TDFAs followed by polarisation controllers. The three signal channels were pre-distorted root-raised-cosine 32~GBaud 64~QAM signals, with a test channel at the centre surrounded by two neighbouring dummy channels with 32.5~GHz channel spacing. Co-propagating WDM channels were emulated by spectrally-shaped ASE noise generated by a wideband ASE source and shaped by WaveShapers\textsuperscript{\textregistered} which operated as wavelength selective switches (WSSs). The recirculating loop consisted of a pair of acousto-optic modulators to switch between loading and recirculating states, a polarisation scrambler (PS) to avoid the accumulation of polarisation-dependent losses, three gain blocks (each of which included lumped repeaters: S-TDFA, low-gain C-EDFA and low-gain L-EDFA), a 71-km low OH peak SSMF, four backward Raman pumps, and WSSs and variable optical attenuators (VOAs) for balancing the loop powers. The Raman pump wavelengths and powers are listed in Table~\ref{tab:pump}. After 1065~km transmission, the signal was directed to the receiver block, which consisted of a bandpass filter, a pre-amplifier, a 90-degree hybrid for coherent detection, and a 256~GSa/s real-time oscilloscope. Offline pilot-based digital signal processing~\cite{wakayama2021_2048} was based on 2 samples per symbol with pilot insertion rate of 1/32, and the SNR was evaluated for all WDM channels. The total throughput, estimated from the generalised mutual information of each channel, was 121.27~Tb/s over 5 spans (365~km), and 94.28~Tb/s over 15 spans (1065~km).  

\begin{table}\footnotesize
\centering
    \caption{Raman pump configuration.}
    \label{tab:pump}
    \begin{tabular}{c|c|c|c|c|c}
        \hline
        $\lambda$ (nm) & 1365 & 1385 & 1405 & 1425 & Total\\
        \hline
        P (mW) & 505.8 & 374.1 & 324.3 & 295.8 & 1500.0\\
        \hline
    \end{tabular}
    \vspace{-17pt}
\end{table}

\begin{figure}
    \vspace{-.7em}
    \centering
        \begin{subfigure}{\linewidth}
        \begin{tikzpicture}\footnotesize
            \begin{axis}[
                width=\linewidth,
                height=4cm,
                xlabel={Wavelength (nm)},
                ylabel={Gain / Att. (dB)},
                xmin=1485, xmax=1605,
                ymin=-8, ymax=30,
                xtick={1485,1515,1545,1575,1605},
                xticklabels={1485,1515,1545,1575,1605},
                ytick={0,10,20,30},
                yticklabels={0,10,20,30},
                grid=both,
                xticklabel style={/pgf/number format/1000 sep=},
                xlabel near ticks,
                y label style={at={(axis description cs:.1,.5)}},
                legend pos=south east,
                legend columns=2,
                legend style={xshift=0.6cm,font=\fontsize{6}{0}\selectfont, row sep=-2.5pt,fill=white},
                legend cell align={left},
                axis x line*=bottom,
                clip=false,
            ]
            \addlegendentry{Gain block 1};
            \addlegendimage{Set1-B, line width=1pt}
            \addplot[Set1-B, mark=*, only marks, mark size=0.5, forget plot] table[x=wavelength,y=gain_1]{Loop_Gain_Loss.txt};
            \addlegendentry{Gain block 2};
            \addlegendimage{Set1-C, line width=1pt}
            \addplot[Set1-C, mark=*, only marks, mark size=0.5, forget plot] table[x=wavelength,y=gain_2]{Loop_Gain_Loss.txt};
            \addlegendentry{Gain block 3};
            \addlegendimage{Set1-D, line width=1pt}
            \addplot[Set1-D, mark=*, only marks, mark size=0.5, forget plot] table[x=wavelength,y=gain_3]{Loop_Gain_Loss.txt};
            \addlegendentry{WSS att.};
            \addlegendimage{Set1-A, line width=1pt}
            \addplot[Set1-A, mark=*, only marks, mark size=0.5, forget plot] table[x=wavelength,y=loss_wss]{Loop_Gain_Loss.txt};
            \node[xshift=-0.78cm,yshift=0cm] at (axis cs: (1485,30){(a)};
            \draw (axis cs:1532,28) ellipse (0.1cm and 0.15cm);
            \draw (axis cs:1532,16.2) ellipse (0.1cm and 0.58cm);
            \draw[>=stealth, ->](axis cs:1532,25.7)--(axis cs:1526,25.7);
            \node[anchor=east,font={\fontsize{7pt}{12}\selectfont}] at (axis cs: (1526,25.7){WSS attenuation};
            \draw[>=stealth, ->](axis cs:1532,7.1)--(axis cs:1538,7.1);
            \node[anchor=west,font={\fontsize{7pt}{12}\selectfont}] at (axis cs: (1538,7.1){Lumped repeater gain};
            \end{axis}
        \end{tikzpicture}
        \end{subfigure}
        \begin{subfigure}{\linewidth}
        \begin{tikzpicture}\footnotesize
        
            \begin{groupplot}[group style={group size=2 by 1,horizontal sep=1.5cm},height=3.5cm,width=.5\linewidth]
            
            \nextgroupplot[
                xlabel={Wavelength (nm)},
                ylabel={Att. (dB/km)},
                xmin=1360, xmax=1610,
                ymin=0.16, ymax=0.3,
                xtick={1360,1460,1530,1610},
                xticklabels={1360,1460,1530,1610},
                ytick={0.2,0.25,0.3},
                yticklabels={0.2,0.25,0.3},
                grid=both,
                xticklabel style={/pgf/number format/1000 sep=},
                xlabel near ticks,
                y label style={at={(axis description cs:.15,.5)}},
            ]
                \addplot[Set1-B,line width=1.5pt] table[x=wavelength,y=loss]{loss_sterlite.txt};
                \draw[fill=violet,opacity=0.1,draw=none] (axis cs:1360,0.16) rectangle (axis cs:1460,0.3);
                \draw[fill=blue,opacity=0.1,draw=none] (axis cs:1460,0.16) rectangle (axis cs:1530,0.3);
                \draw[fill=yellow,opacity=0.1,draw=none] (axis cs:1530,0.16) rectangle (axis cs:1565,0.3);
                \draw[fill=red,opacity=0.1,draw=none] (axis cs:1565,0.16) rectangle (axis cs:1610,0.3);
                \node[violet,opacity=0.5] at (axis cs: 1415,0.28) {E};
                \node[blue,opacity=0.5] at (axis cs: 1495,0.28) {S};
                \node[yellow,opacity=1] at (axis cs: 1547,0.28) {C};
                \node[red,opacity=0.5] at (axis cs: 1590,0.28) {L};
                \draw[>=stealth, ->,line width=1pt] (axis cs: 1365,0.2)--(axis cs:1365,0.24);
                \draw[>=stealth, ->,line width=1pt] (axis cs: 1385,0.2)--(axis cs:1385,0.24);
                \draw[>=stealth, ->,line width=1pt] (axis cs: 1405,0.2)--(axis cs:1405,0.24);
                \draw[>=stealth, ->,line width=1pt] (axis cs: 1425,0.2)--(axis cs:1425,0.24);
                \node[anchor=north,font=\footnotesize,align=center]at (axis cs: 1425,0.2){R. pumps};

            \nextgroupplot[
                xlabel={Frequency Sep. (THz)},
                ylabel={R. Gain (1/W/km)},
                xmin=0, xmax=26,
                ymin=0, ymax=0.5,
                xtick={0,8,16,24},
                xticklabels={0,8,16,24},
                ytick={0.1,0.3,0.5},
                yticklabels={0.1,0.3,0.5},
                grid=both,
                xlabel near ticks,
                y label style={at={(axis description cs:.2,.5)}},
            ]
                \addplot[Set1-A,line width=1.5pt] table[x=frequency,y=gain]{RGain_sterlite.txt};
            \end{groupplot}
            \node[xshift=-0.8cm,yshift=1cm] at (group c1r1.west){(b)};
            \node[xshift=0.25cm,yshift=.7cm] at (group c2r1.west){(c)};
        \end{tikzpicture}
        \end{subfigure}
    \caption{Wavelength-dependent parameters. (a) Gain of gain block 1, 2, and 3, and loop WSS attenuation. (b) Fibre attenuation. (c) Raman gain profile.}
    \label{fig:para}
\end{figure}
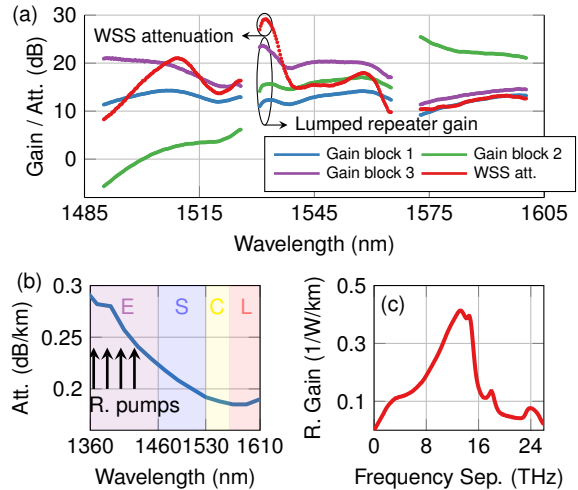

The closed-form ISRS GN model~\cite{buglia2023closedraman} in the presence of Raman amplification was used to estimate the NLI and the ASE noise. The ASE noise calculation consists of two components: one generated in the transmission fibre, obtained from the Raman-coupled equations~\cite{bromage2004raman,buglia2024throughput} and the signal profile evolution~\cite{buglia2023closedraman}, and the other from the lumped amplifiers (gain blocks in Fig.~\ref{fig:exp}). Transceiver noise-limited SNR was measured from the experiment in a B2B configuration, as shown in the green dotted path in Fig.~\ref{fig:exp}. The parameters of the fibre and components are listed in Table~\ref{tab:para}, and the insertion losses of the optical components in the loop were assumed to be wavelength-independent in the model. The wavelength dependence of the repeater amplifiers gain, loop WSS shaping loss, fibre attenuation and Raman gain are shown in Fig.~\ref{fig:para}. The aforementioned parameters were used in the model to accurately reproduce the experimental setup to calculate the NLI and ASE noise.

\begin{table}\footnotesize
\centering
    \caption{System and fibre parameters considered in the model.}
    \label{tab:para}
    \begin{tabular}{c|c|c}
        \hline
        Parameter & Unit & Value\\
        \hline
        S/C/L amplifier noise figure & dB & 7.5/5/6\\
        \hline
        Nonlinear parameter $\gamma$ & $\text{W}^{-1}\text{km}^{-1}$ & 1.4\\
        \hline
        Effective core area $A_{\text{eff}}$ & $\mu\text{m}^{2}$ & 83\\
        \hline
        Dispersion $D$ at 1550~nm& $\text{ps}\cdot\text{nm}^{-1}\text{km}^{-1}$ & 16.5\\
        \hline
        Dispersion slope $S$ & $\text{ps}\cdot\text{nm}^{-2}\text{km}^{-1}$ & 0.09\\
        \hline
    \end{tabular}
\end{table}

\vspace{-8pt}
\section{Results and Discussion}
As shown in Fig.~\ref{fig:para}(a), gain block~1 amplifies the signal by approximately 13~dB to overcome the insertion loss of AOMs, couplers, and the PS, followed by VOAs to control the launch powers in each band. Gain block~2 compensates for the fibre loss. Since Raman pumps mainly provide gain in the S-band and part of the C-band, gain block~2 gave the most gain in L-band with more than 20~dB, contributing to most of the ASE noise. In contrast, S-band had the lowest gain from lumped amplification, and the negative gain at the shortest wavelength means that the Raman pump recovered the signal power to a level higher than the launch power. The C-band behaviour was in the middle, with an average gain of 15~dB. Gain block~3 compensated for the WSS insertion losses of 4.5~dB in S-band and 7~dB in C-, L-bands, and additional shaping loss. The WSS attenuation in Fig.~\ref{fig:para}(a) approximately reflects the Raman on/off gain, with a peak in the gain at a wavelength of 1505~nm. The other peak at the shortest C-band wavelengths is due to the gain from the 1425~nm pump. 

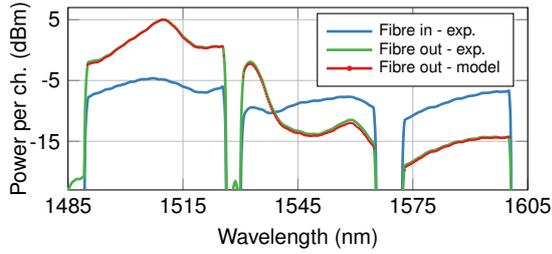
\begin{figure}[t!]
    \centering
        \begin{tikzpicture}\footnotesize
            \begin{axis}[
                width=\linewidth,
                height=4cm,
                xlabel={Wavelength (nm)},
                ylabel={Power per ch. (dBm)},
                xmin=1485, xmax=1605,
                ymin=-23, ymax=7,
                xtick={1485,1515,1545,1575,1605},
                xticklabels={1485,1515,1545,1575,1605},
                ytick={-15,-5,5},
                yticklabels={-15,-5,5},
                grid=both,
                xticklabel style={/pgf/number format/1000 sep=},
                xlabel near ticks,
                y label style={at={(axis description cs:.1,.5)}},
                legend pos=north east,
                legend columns=1,
                legend style={font=\fontsize{6}{0}\selectfont, row sep=-2.5pt,fill=white},
                legend cell align={left},
            ]
            \addlegendentry{Fibre in - exp.};
            \addplot[Set1-B, line width=1pt] table[x=wavelength,y=power_in]{Spectrum_Fibre_In_Out.txt};
            \addlegendentry{Fibre out - exp.};
            \addplot[Set1-C, line width=1pt] table[x=wavelength,y=power_out]{Spectrum_Fibre_In_Out.txt};
            \addlegendentry{Fibre out - model};
            \addlegendimage{Set1-A, line width=1pt, mark=*, mark size=0.5pt}
            \addplot[Set1-A, only marks, mark=*, mark size=0.2pt, forget plot] table[x=wavelength,y=model_out]{Spectrum_Fibre_Out_Model.txt};
            \end{axis}
        \end{tikzpicture}
    \caption{Signal power spectra measured in the experiment (blue and green solid line) and predicted by the model (red dots).}
    \label{fig:spec}
\end{figure}

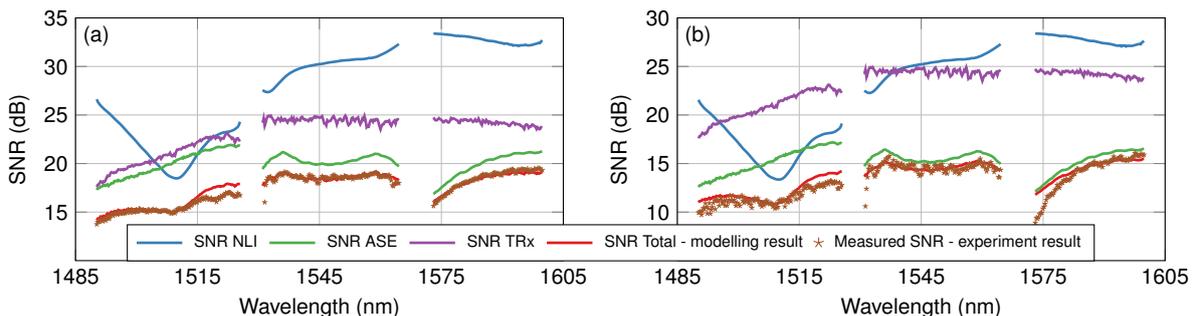
\begin{figure*}[b!]
    \vspace{-1em}
    \centering
        \begin{tikzpicture}\footnotesize
            \begin{groupplot}[group style={group size=2 by 1,horizontal sep=1.5cm},height=4.8cm,width=8cm]
            \nextgroupplot[
                xlabel={Wavelength (nm)},
                ylabel={SNR (dB)},
                xmin=1485, xmax=1605,
                ymin=10, ymax=35,
                xtick={1485,1515,1545,1575,1605},
                xticklabels={1485,1515,1545,1575,1605},
                ytick={15,20,25,30,35},
                yticklabels={15,20,25,30,35},
                grid=both,
                xticklabel style={/pgf/number format/1000 sep=},
                xlabel near ticks,
                ylabel near ticks,
                legend columns=2,
                legend style={xshift=0.6cm,font=\fontsize{6}{0}\selectfont, row sep=-2.5pt,at={(axis description cs:1.2,0.24)},fill=white},
                legend cell align={left},
            ]
            \addplot[Set1-B,line width=1pt] table[x=wav,y=NLI]{model_data_5spans_s.txt};
            \addplot[Set1-B,line width=1pt] table[x=wav,y=NLI,forget plot]{model_data_5spans_c.txt};
            \addplot[Set1-B,line width=1pt] table[x=wav,y=NLI,forget plot]{model_data_5spans_l.txt};
            \addplot[Set1-C,line width=1pt] table[x=wav,y=ASE]{model_data_5spans_s.txt};
            \addplot[Set1-C,line width=1pt] table[x=wav,y=ASE,forget plot]{model_data_5spans_c.txt};
            \addplot[Set1-C,line width=1pt] table[x=wav,y=ASE,forget plot]{model_data_5spans_l.txt};
            \addplot[Set1-D,line width=1pt] table[x=wav,y=B2B]{model_data_5spans_s.txt};
            \addplot[Set1-D,line width=1pt] table[x=wav,y=B2B,forget plot]{model_data_5spans_c.txt};
            \addplot[Set1-D,line width=1pt] table[x=wav,y=B2B,forget plot]{model_data_5spans_l.txt};
            \addplot[Set1-A,line width=1pt] table[x=wav,y=Total,forget plot]{model_data_5spans_s.txt};
            \addplot[Set1-A,line width=1pt] table[x=wav,y=Total,forget plot]{model_data_5spans_c.txt};
            \addplot[Set1-A,line width=1pt] table[x=wav,y=Total,forget plot]{model_data_5spans_l.txt};
            \addplot[Set1-G,only marks,mark=star,opacity=1,mark size=1] table[x=wav,y=Exp]{model_data_5spans_s.txt};
            \addplot[Set1-G,only marks,mark=star,opacity=1,mark size=1] table[x=wav,y=Exp]{model_data_5spans_c.txt};
            \addplot[Set1-G,only marks,mark=star,opacity=1,mark size=1] table[x=wav,y=Exp]{model_data_5spans_l.txt};
            \node[anchor=north west] at (axis cs: (1485,35){(a)};
            
            \nextgroupplot[
                xlabel={Wavelength (nm)},
                ylabel={SNR (dB)},
                xmin=1485, xmax=1605,
                ymin=5, ymax=30,
                xtick={1485,1515,1545,1575,1605},
                xticklabels={1485,1515,1545,1575,1605},
                ytick={10,15,20,25,30},
                yticklabels={10,15,20,25,30},
                grid=both,
                xticklabel style={/pgf/number format/1000 sep=},
                xlabel near ticks,
                ylabel near ticks,
                legend columns=5,
                legend style={xshift=1cm,font=\fontsize{6}{0}\selectfont, column sep=2pt,at={(axis description cs:0.7,0.15)},fill=white},
                legend cell align={left},
            ]
            \addlegendentry{SNR NLI}; 
            \addplot[Set1-B,line width=1pt] table[x=wav,y=NLI]{model_data_15spans_s.txt};
            \addplot[Set1-B,line width=1pt] table[x=wav,y=NLI,forget plot]{model_data_15spans_c.txt};
            \addplot[Set1-B,line width=1pt] table[x=wav,y=NLI,forget plot]{model_data_15spans_l.txt};
            \addlegendentry{SNR ASE};
            \addplot[Set1-C,line width=1pt] table[x=wav,y=ASE]{model_data_15spans_s.txt};
            \addplot[Set1-C,line width=1pt] table[x=wav,y=ASE,forget plot]{model_data_15spans_c.txt};
            \addplot[Set1-C,line width=1pt] table[x=wav,y=ASE,forget plot]{model_data_15spans_l.txt};
            \addlegendentry{SNR TRx};
            \addplot[Set1-D,line width=1pt] table[x=wav,y=B2B]{model_data_15spans_s.txt};
            \addplot[Set1-D,line width=1pt] table[x=wav,y=B2B,forget plot]{model_data_15spans_c.txt};
            \addplot[Set1-D,line width=1pt] table[x=wav,y=B2B,forget plot]{model_data_15spans_l.txt};
            \addlegendentry{SNR Total - modelling result};
            \addplot[Set1-A,line width=1pt] table[x=wav,y=Total]{model_data_15spans_s.txt};
            \addplot[Set1-A,line width=1pt] table[x=wav,y=Total,forget plot]{model_data_15spans_c.txt};
            \addplot[Set1-A,line width=1pt] table[x=wav,y=Total,forget plot]{model_data_15spans_l.txt};
            \addlegendentry{Measured SNR - experiment result};
            \addlegendimage{only marks, mark=star, mark size=2, mark options={opacity=1}, mark options={solid,draw=Set1-G}}
            \addplot[Set1-G,only marks,mark=star,opacity=1,mark size=1] table[x=wav,y=Exp]{model_data_15spans_s.txt};
            \addplot[Set1-G,only marks,mark=star,opacity=1,mark size=1] table[x=wav,y=Exp]{model_data_15spans_c.txt};
            \addplot[Set1-G,only marks,mark=star,opacity=1,mark size=1] table[x=wav,y=Exp]{model_data_15spans_l.txt};
            \node[anchor=north west] at (axis cs: (1485,30){(b)};
            \end{groupplot}
        \end{tikzpicture}
    \caption{SNR limited by the transceiver, ASE, NLI, and total SNR estimated from the model and measured from experiment. Solid lines: model prediction. Star markers: experimental results. (a) 5 recirculations (355~km). (b) 15 recirculations (1065~km).}
    \label{fig:SNR}
\end{figure*}

Figure~\ref{fig:spec} shows the fibre launch spectrum with 18.3~dBm, 15.1~dBm, and 14.8~dBm in the S-, C\nobreakdash-, and L-band, respectively. The C- and L-band launch power per channel was set to approximately 2~dB below that of the S-band to reduce the ISRS power transfer, such that the S-band performance would not be fully dominated by ASE noise but would be affected by ASE and NLI noise jointly.
The fibre output signal spectrum (green solid line) is shown in Fig.~\ref{fig:spec}, together with the predicted output from the Raman coupled equation (red dots) in~\cite{buglia2024throughput}. The model accurately estimated the signal power profile evolution over the 71~km fibre link in the presence of backward Raman amplification, with an average error of only 0.23~dB. This power evolution was further used to calculate the Raman-coupled ASE noise in the fibre link.

Figure~\ref{fig:SNR} shows the various noise contributions to the total SNR in 5-span and 15-span transmission, calculated using the closed-form GN model. The maximum NLI noise is observed at 1505~nm due to the high gain from the Raman amplification, while lower NLI is seen in C- and L-bands due to the lower launch power. The ASE noise level is similar in the three bands; in the case of the S-band, the higher noise figure of the TDFA is offset by the better noise performance of the DRA. A clear dip in total SNR due to the NLI at around 1505~nm was accurately predicted in both cases by the model, confirming the accuracy of the NLI noise predicted by the model. Comparing the experimental results (star markers) and the model predictions (red solid line) in Fig.~\ref{fig:SNR}, the average SNR estimation errors are 0.38~dB and 0.60~dB for 5- and 15-span transmission, respectively. The discrepancy in the longest wavelength S-band and the shortest wavelength L-band is due to the wavelength dependence of amplifier noise figures and the insertion losses of optical components in the loop, which were assumed to be wavelength-independent in the model. These results experimentally validate the recently proposed GN model in the presence of Raman amplification, confirming its suitability as a key tool in optimising the system design and maximising the throughput of future hybrid-amplified long-haul UWB transmission systems.

\section{Conclusions}
We demonstrated the powerful capability for accurate SNR estimation in long-haul WDM transmission using a recently-developed closed-form GN model in the presence of distributed Raman amplification. The S+C+L-band recirculating fibre loop WDM transmission experiment, with hybrid Raman-EDFA-TDFA amplification, over standard single mode fibre link lengths of 365~km and 1065~km, showed a discrepancy between the experiments and the model of less than 0.23~dB in signal power spectrum evolution, and of only 0.38~dB and 0.60~dB, respectively in the SNR, averaged across all channels.

\section{Acknowledgements}
This work was supported by EPSRC grants EP/R035342/1 TRANSNET (Transforming networks - building an intelligent optical infrastructure), EP/W015714/1 EWOC (Extremely Wideband Optical Fibre Communication Systems), EP/V000969/1 ARGON (All-Raman optical amplification for next Generation ultra-wideband Optical Networks), EP/V007734/1 EPSRC Strategic equipment grant, EP/T517793/1 EPSRC studentship, and Microsoft Research.

\printbibliography

\vspace{-4mm}

\end{document}